\documentclass{webofc}

\usepackage[varg]{txfonts}   
\usepackage{hyperref}
\usepackage{url}
\hypersetup{colorlinks=true,citecolor=blue,urlcolor=blue,linkcolor=blue}
\usepackage{subcaption}

\newcommand{\Gq}[6]{\mathcal{G}^{#1}(#2,\boldsymbol{#3}; #4, \boldsymbol{#5}| #6)}

\newcommand{\Exp}[1]{\exp{\left\{#1\right\}}}

\newcommand{\nn}{\nonumber\\ }

\def\P{{\boldsymbol P}}

\def\p{{\boldsymbol p}}

\def\x{{\boldsymbol x}}

\def\r{{\boldsymbol r}}

\def\a{{\boldsymbol a}}

\def\Re{\text{Re} \, }

\newcommand{\cK}{\mathcal{K}}

\newcommand{\cA}{\mathcal{A}}

\begin{document}
\title{Quark antenna in early stage anisotropic QCD matter}
%
%

\author{\firstname{João} \lastname{Barata}\inst{1}\fnsep\thanks{\email{joao.lourenco.henriques.barata@cern.ch}} \and
        \firstname{Carlos} \lastname{A. Salgado}\inst{2,3}\fnsep\thanks{\email{carlos.salgado@usc.es}} \and
        \firstname{João} \lastname{M. Silva}\inst{4,5}\fnsep\thanks{\email{joao.m.da.silva@tecnico.ulisboa.pt}}
}

\institute{CERN, Theoretical Physics Department, CH-1211 Geneva 23, Switzerland 
\and
    Instituto Galego de Física de Altas Enerxías IGFAE, Universidade de Santiago de Compostela, E- 15782 Galicia-Spain         
\and
    Axencia Galega de Innovación (GAIN), Xunta de Galicia, Galicia-Spain
\and 
    Laboratório de Instrumentação e Física Experimental de Partículas (LIP), Av. Prof. Gama Pinto, 2, 1649-003 Lisbon, Portugal      
\and 
    Departamento de Física, Instituto Superior Técnico (IST), Universidade de Lisboa, Av. Rovisco Pais 1, 1049-001 Lisbon, Portugal
          }

\abstract{The states of matter produced in the early stage of heavy ion collisions can be highly anisotropic. If such a feature is sufficiently pronounced, one should expect the final particle distribution inside jets to reflect it in the form of non-trivial angle correlations. In this talk, we discuss a first step in exploring such correlations by studying how a $q\bar q$ state produced from an initial unpolarized gluon couples to the anisotropies of an underlying static QCD medium. The medium anisotropy is captured by allowing the jet quenching parameter to take different magnitudes in two orthogonal directions in the plane transverse to the jet axis.
We find that the final particle distribution is sensitive to the medium anisotropy in the form of an azimuthal angle modulation, and more importantly, that this effect couples directly to the helicity/spin of the final states, offering a novel way to extract the details of the underlying matter which is not accessible with standard jet observables. We further show how such features can be extracted from the Fourier decomposition of the distribution and from final state transverse spin polarization measurements.
}
\maketitle
\section{Introduction}
\label{sec:intro}

In ultra-relativistic heavy-ion collisions (URHICs) at RHIC and at the LHC, it is generally believed that, in the first $1$ fm/c (or $3$ ys) after the ions collide, the produced matter is in a highly anisotropic and out-of-equilibrium state, dominated by strong polarized chromoelectric and magnetic fields, which eventually evolve into a close to equilibrium state preceding the quark-gluon plasma (QGP) phase~\cite{Schlichting:2019abc,Berges:2020fwq}. To probe such QCD states of matter, and ultimately extract the transport properties of the QGP, one can leverage the production of hadronic jets. The study of medium-induced jet modifications (cf.  ~\cite{Qin:2015srf, Cunqueiro:2021wls, Apolinario:2022vzg} for reviews), is typically performed using simplified medium models, mainly focused on capturing the equilibrated QGP phase, commonly assuming the medium to be static, isotropic and homogeneous. Although such assumptions allow for interesting studies of in-medium jets (see e.g.~\cite{Casalderrey-Solana:2012evi,Mehtar-Tani:2010ebp, Apolinario:2024apr}), the states of matter produced in real URHICs, as mentioned above, can be highly dynamical and anisotropic. If such type of effects are sufficiently strong, one should expect the final particle distribution inside jets to reflect an angular \textit{modulation}, which can be correlated to the medium's structure.\footnote{We note nonetheless that since the hydrodynamic expansion tends to isotropize the system, many of these effects can be ultimately washed out.}

In this work, we take a step in this direction by studying how a $q\bar q$ pair branching from an initial unpolarized gluon couples to the anisotropies of an underlying QCD medium, similarly to what was done recently for pure gluonic systems~\cite{Hauksson:2023tze}. Assuming the medium to be sufficiently long, we compute the leading order cross-section for this process in the large $N_c$ limit, and show that the final distribution has a non-trivial azimuthal profile which depends on the final states' helicity/spin. The most important element in our calculation, following~\cite{Hauksson:2023tze}, is that we allow the jet quenching parameter $\hat q$ to take different values along two orthogonal directions in the transverse plane to the jet axis, which we shall denote as $x$ and $y$.\footnote{In what follows we shall not discuss the physics from which the anisotropic $\hat q$ emerges. Equivalent forms for this parameter have been found in the early stages of heavy ions collisions~\cite{Boguslavski:2023waw,Boguslavski:2023alu,Ipp:2020mjc,Ipp:2020nfu,Avramescu:2023qvv,Carrington:2021dvw,Carrington:2022bnv,Barata:2024xwy}, and in the QGP epoch~\cite{Kuzmin:2023hko,Barata:2023zqg,Barata:2023qds,Barata:2022utc,Andres:2022ndd,Barata:2022krd,Sadofyev:2021ohn,He:2020iow,Xiao:2024ffk,Fu:2022idl,He:2022evt,Kuzmin:2024smy}. See also~\cite{Hauksson:2020wsm,Hauksson:2021okc} for related work.} The observables proposed in this work would contribute to pin down the relevant dynamics associated to an anisotropic transport coefficient.

\section{Quark pair production in a dense anisotropic QCD medium }
\label{sec:calculation}
We start by deriving the double differential cross-section for the $g\rightarrow q\bar q$ process (i.e. a quark antenna) inside a medium of length $L^+$ which we assume is large enough, such that we fix the branching to happen inside the medium. In the limit of a dense QCD background medium, one can calculate this in a path integral formalism, as originally discussed by BDMPS-Z~\cite{Zakharov:1996fv,Baier:1996kr} (see e.g.~\cite{Mehtar-Tani:2013pia,Blaizot:2015lma} for recent reviews and~\cite{Attems:2022ubu} for a recent related calculation). In this approach, to each line in the partonic process one associates an effective quantum mechanical propagator, which for massless quarks and in the light-cone gauge takes the form
\begin{equation}
    \Gq{ij}{y^+}{y}{x^+}{x}{p^+} = \int^{\boldsymbol{r}(y^+) = \boldsymbol{y}}_{\boldsymbol{r}(x^+)=\boldsymbol{x}}\mathcal{D}\boldsymbol{r}(\xi)\Exp{i\frac{p^+}{2}\int_{x^+}^{y^+}ds^+ \, \dot{\boldsymbol{r}}^2}U^{ij}(x^+,y^+,\boldsymbol{r}(\xi)) \, ,
\end{equation}
with the endpoints denoting the starting transverse positions of the quark evolution in light-cone time $x^+$, at fixed light-cone energy $p^+$. We assume $p^+ \gg |\p|$ to be the largest scale in the problem, such that the evolution is highly collimated and all the dynamics are constrained to the plane transverse to the jet axis. The successive scatterings with the medium lead to color precession, which is captured by the path-ordered fundamental Wilson line
\begin{equation}
    U^{ij}(x^+,y^+,\boldsymbol{r}) = \mathcal{P}\Exp{ig\int_{x^+}^{y^+}ds^+\, \cA_a^{-}(s^+, \boldsymbol{r}(s^+))t^a_{ij}}\, .
\end{equation}
Here $\mathcal{A}$ denotes the stochastic background field representing the underlying QCD medium. When computing any observable one should average over field configurations, which we assume follow a Gaussian distribution
\begin{equation}\label{eq:pair_correlator} 
    \langle \mathcal{A}_a^-(x^+,x^-,\boldsymbol{x})\mathcal{A}_b^-(y^+,y^-,\boldsymbol{y})\rangle = \delta_{ab}n(x^+)\delta(x^+-y^+)\gamma(\boldsymbol{x-y})\, ,
\end{equation}
The function $\gamma(\x)$ is directly related to the in-medium elastic scattering rate and $n(s^+)$ is the density of the medium which, in this work, takes the form $n_0\, \Theta(s^+ < L^+)$, i.e., a static medium of length $L^+$, which is usually taken to be of the order of a few fm. Applying effective Feynman rules (see e.g.~\cite{Barata:2024bqp}), one can compute the cross-section for  $g\rightarrow q\bar q$, resumming all possible gluon exchanges with the medium at leading order in eikonality.
The medium average involved in this computation can be greatly simplified since Eq.~\eqref{eq:pair_correlator} is local in color and light-cone time, allowing one to write it as a product $\mathcal{S}^{(2)}\mathcal{S}^{(3)}\mathcal{S}^{(4)}$, corresponding to two, three and four body averages. They describe, respectively, gluon broadening, the production of the $q\bar q$ antenna and the independent broadening of the quark and anti-quark. The latter is only simplified analytically as such in the large-$N_c$ limit. Further, in this limit, the resulting path integrals can all be simplified analytically, resulting in a splitting \textit{kernel} of the form
\begin{align}\label{eq:general_kernel}
    \cK(\a_1, \a_2) = \int_{\a_1}^{\a_2}\mathcal{D}\boldsymbol{u}\Exp{\frac{i\tilde{q}_0^+}{2}\int_{x_1^+}^{x_2^+} ds^+\left(\dot{\boldsymbol{u}}^2+\frac{i\, n(s^+) N_c\left(\sigma(\boldsymbol{u}) + \sigma((1-z)\boldsymbol{u})\right)}{2\tilde q_0^+}\right)} \, ,
\end{align}
where $\tilde{q}_0^+=z(1-z)q_0^+$, $q_0^+$ the light-cone energy of the initial gluon, $z$ the energy fraction of the quark and the dipole cross-section is defined as $\sigma(\r) = 2g^2(\gamma(0)-\gamma(\r))$.  To solve Eq.~\eqref{eq:general_kernel}, we use the harmonic approximation, i.e., we drop the logarithmic dependence on the transverse position, in the small distance limit
\begin{equation}
        N_c \int_{x^+}^{y^+} ds^+ \, n(s^+) \sigma(\boldsymbol{r}) = N_c  (y^+-x^+) n \sigma(\boldsymbol{r}) = \frac{(y^+-x^+)}{2}\left(\hat{q}_x r_x^2 + \hat{q}_y r_y^2\right)  + \mathcal{O}(\r^2\log\r^2)\, ,
\end{equation}
where we have assumed the medium to be static and we wrote the components of the transverse vector as $\r = (r_x,r_y)$. A form of medium anisotropy is naturally introduced here by considering different values for the jet quenching parameter $\hat q$ in different directions in the transverse plane \cite{Hauksson:2023tze}. After integrating out the total transverse momentum of the system ($\p_1 + \p_2$), we write the spectrum differential in $z$ and on a single angle $\phi$ corresponding to the azimuth of the relative transverse momentum of the $q\bar q$ pair $\P^{\rm rel} = (1-z)\p_1 -z\p_2$, i.e., $\tan\phi = \P^{\rm rel}_y/\P^{\rm rel}_x$.
The final analytical result for the in-medium $g\rightarrow q\bar q$ spectrum then reads
\begin{align}\label{massive_zaligned_final}
    & \frac{dN^{hh'}}{dzd\phi} \equiv \frac{1}{\sigma_0}\frac{d\sigma^{hh'}}{dzd\phi} = T_F\frac{\alpha_s}{(2\pi)^2} \Re  \int_{0<x_v^+ < \bar x_v^+ < L} \frac{\sqrt{c_{1x}}\sqrt{c_{1y}}}{\sqrt{c_{3x}}\sqrt{c_{3y}}} e^{-i\frac{m^2}{2\tilde{q}_0^+}\Delta t}\nn
    & \times \Bigg\{\delta_{hh'}\left(P_{qg}^{\rm vac}(z)\left[\frac{(c_{1y}c_{2y}c_{3x}-c_{1x}c_{2x}c_{3y})(c_{3y}\cos^2{\phi}-c_{3x}\sin^2{\phi})}{\left(c_{3y}\cos^2{\phi}+c_{3x}\sin^2{\phi}\right)^2}+\frac{c_{3x}c_{3y}c_4}{c_{3y}\cos^2{\phi}+c_{3x}\sin^2{\phi}}\right]\right.\nn
    & \left.  + 2ih(1-2z) \frac{ c_{3x}c_{3y}c_5\sin{\phi}\cos{\phi}}{\left(c_{3y}\cos^2{\phi}+c_{3x}\sin^2{\phi}\right)^2}\right) + \delta_{h,-h'}\left(\frac{m^2}{2\tilde{q}_0^+}\frac{c_{3x}c_{3y}}{c_{3y}\cos^2{\phi}+c_{3x}\sin^2{\phi}}\right)\Bigg\} \, ,
\end{align}
where we have introduced the compact notation
\begin{align}\label{eq:ci_definition}
    &c_{1i} = \frac{\Omega_i}{2i\sin{\Omega_i\Delta t}}\,, & & c_{2i} = \frac{\Omega_i}{\tan{\Omega_i\Delta t}}\,,& & 
    c_{3i} = \Delta_L^+\Omega_i^2-c_{2i}\, , \nn
    & c_4 = \left(c_{1x}+c_{1y}\right)\, ,& & c_5 = \left(c_{1y}c_{2x}-c_{1x}c_{2y}\right) \,, & & \Omega = \frac{1-i}{\sqrt{2}}\sqrt{\frac{\hat q_i P^{\rm vac}_{qg}(z)}{4\tilde q_0^+}}
\end{align}
and $P^{\rm vac}_{qg}(z) = z^2 + (1-z)^2$ is the $g\rightarrow q\bar q$ massless vacuum splitting function stripped of its color factor ($T_F = 1/2$). The integrated initial production cross-section is given by $\sigma_0$.

\section{Jet observables}
\label{sec:observables}
The following numerical results are presented as a function of the dimensionless quantities
\begin{equation}\label{dimensionless_var_def}
    \zeta = \frac{\sqrt{\hat q_y} - \sqrt{\hat q_x}}{\sqrt{\hat q_y} + \sqrt{\hat q_x}}\, ,\qquad  r = L^+ \frac{\left(\sqrt{\hat q_y} + \sqrt{\hat q_x}\right)}{2\sqrt{2q_0^+}} \, , \qquad \mu = \frac{2 m^2}{\left(\sqrt{\hat q_y} + \sqrt{\hat q_x}\right)\sqrt{q_0^+}} \, ,
\end{equation}
The values for $r \sim 2$ and $\mu \sim 0.16$ used in the following plots can be estimated by considering $L^+/\sqrt{2} \sim 5$ fm, $q_0^+/\sqrt{2} \sim 100$ GeV, $(\sqrt{\hat q_x} + \sqrt{\hat q_y})/2 \sim 2 \,\, {\rm GeV}\,\, {\rm fm}^{-1/2}$ and the charm quark mass $m = 1.27 $ GeV/$c^2$.

\begin{figure}
\centering
\begin{subfigure}[h]{0.49\textwidth}
         \centering
         \includegraphics[width=\textwidth]{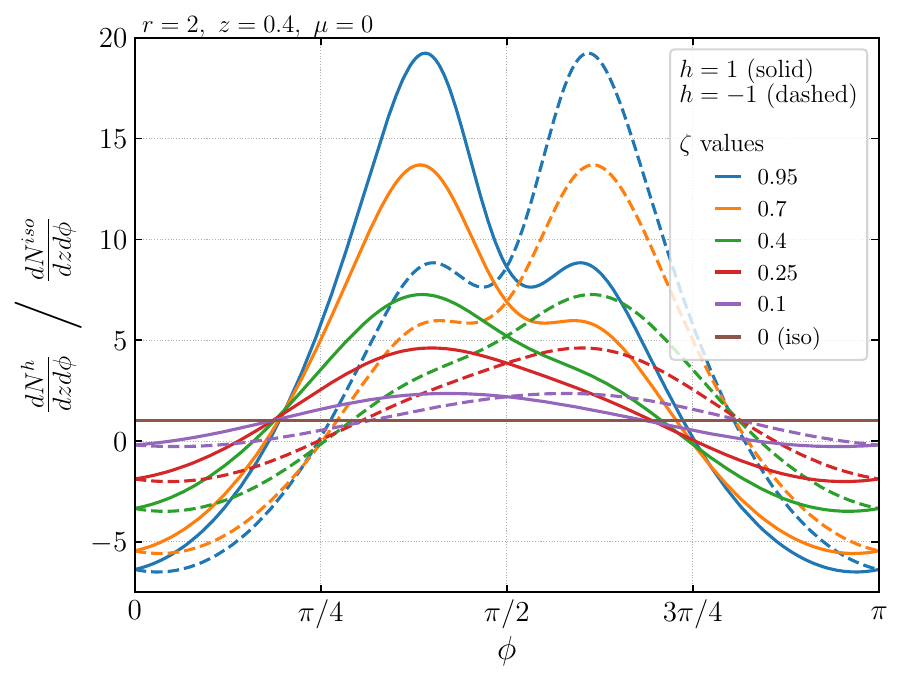}
     \end{subfigure}
     \begin{subfigure}[h]{0.49\textwidth}
         \centering
         \includegraphics[width=\textwidth]{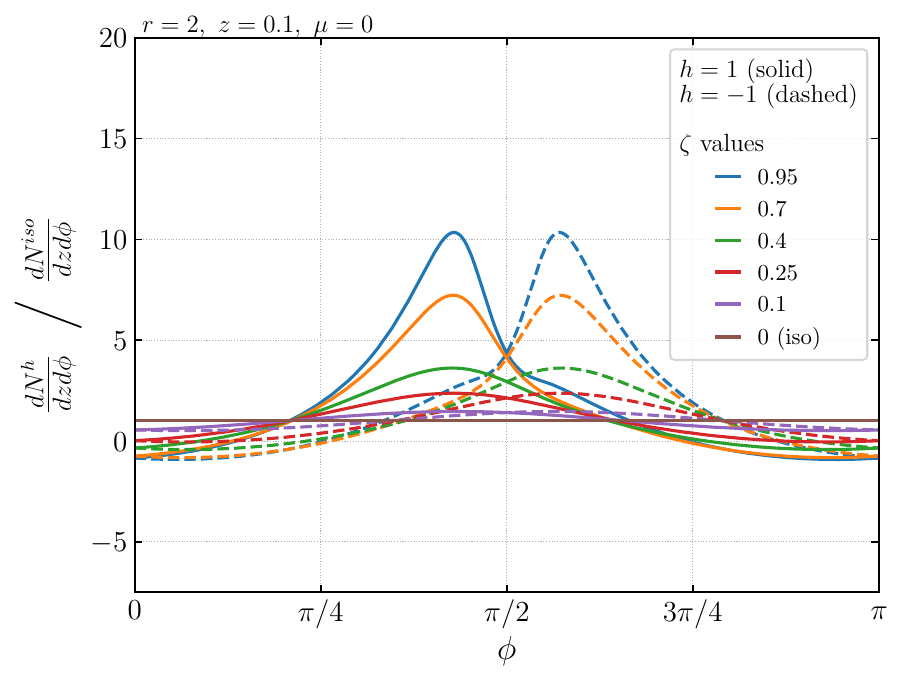}
     \end{subfigure}
\caption{Ratio of the massless ($\mu=0$) azimuthal distribution (Eq.~\eqref{massive_zaligned_final}) with respect to the isotropic case ($\zeta=0$) as a function of $\phi$ for different values of the anisotropy parameter $\zeta$, for fixed $r=2$, for more energetically balanced splittings ($z=0.4$, \textbf{left}) and more unbalanced splittings ($z=0.1$, \textbf{right}). The solid curve corresponds to helicity state $h=1$ and the dashed curve to $h=-1$. The vacuum piece is subtracted.}
\label{azimuthal_plot_phi}       
\end{figure}

\textbf{Azimuthal distribution:} In Fig.~\ref{azimuthal_plot_phi} we show the distribution in Eq.~\eqref{massive_zaligned_final} for massless quarks as a function of $\phi$, for several values of the anisotropic coefficient $\zeta > 0$ ($\zeta <0$ just switches the axes) and for two values of energy fraction -- $z=0.4$ (left panel) and $z=0.1$ (right panel). The plots show the ratio between the azimuthal distribution for a given helicity state ($h=1$ solid, $h=-1$ dashed) and the isotropic case ($\zeta=0$). The isotropic distribution is flat ($\phi$ independent). We see that a larger anisotropy results in a larger ratio near $\pi/2$, where the distinction with respect to the isotropic case seems the largest. The ratio is not the largest exactly at $\pi/2$ because there is helicity dependence. We further observe that for small anisotropies the distributions become increasingly simple and more symmetric with respect to $\phi=\pi/2$, which indicates that the helicity independent terms (symmetric under $\phi\rightarrow \pi-\phi$) dominate over the dependent ones (anti-symmetric under $\phi\rightarrow \pi-\phi$). Comparing the solid and dashed curves of each color, one concludes that one can distinguish the two helicity states better the greater the anisotropy and, seemingly, the softest the splitting (owing to the $(1-2z)$ factor) and this is again more pronounced near $\pi/2$.

\begin{figure}
\centering
\begin{subfigure}[h]{0.49\textwidth}
         \centering
         \includegraphics[width=\textwidth]{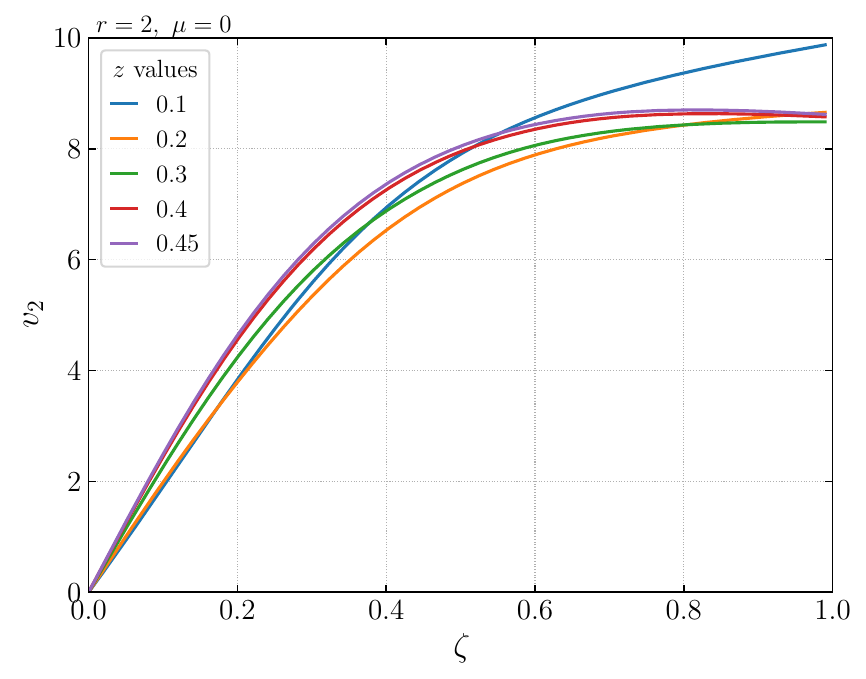}
     \end{subfigure}
     \begin{subfigure}[h]{0.49\textwidth}
         \centering
         \includegraphics[width=\textwidth]{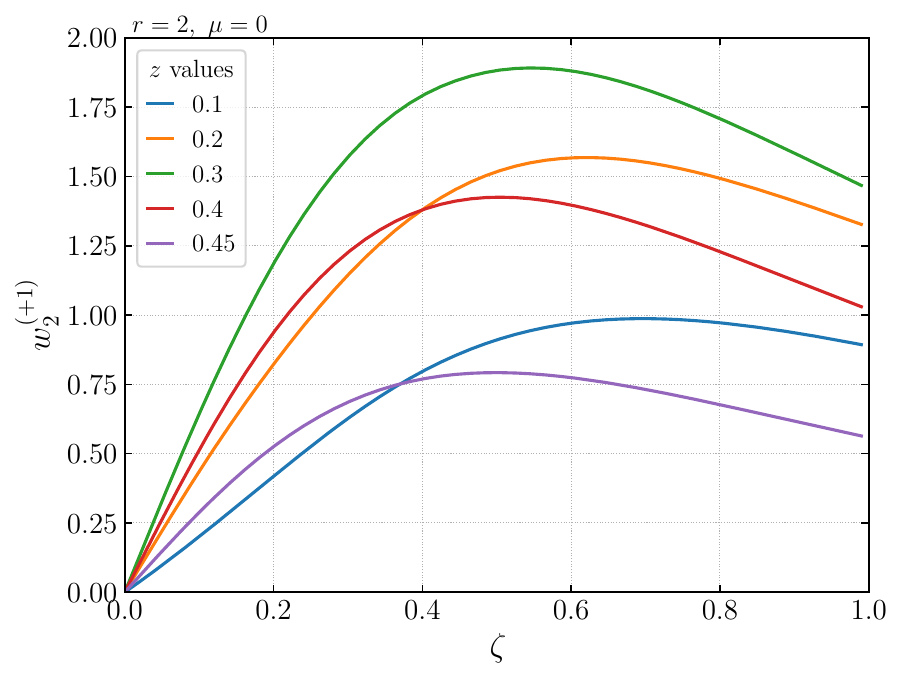}
     \end{subfigure}
\caption{\textbf{Left:} Evolution of  $v_2$ in Eq.~\eqref{eq:harmonic_exp} as a function of $\zeta$ and for different values of $z$, with fixed $r=2$ and for massless quarks $(\mu=0)$. \textbf{Right:} equivalent plots for $w_2^{(+1)}$ in Eq.~\eqref{eq:harmonic_exp}. The vacuum piece is subtracted.}
\label{fig:massless_harm}       
\end{figure}

\textbf{Fourier decomposition:} The symmetry properties showcased in the previous plots suggest that one can isolate the helicity dependent terms through a harmonic Fourier decomposition of the form: 
\begin{align}\label{eq:harmonic_exp}
   \frac{2\pi}{ (dN/dz)} \frac{dN^{h}}{dzd\phi}  = 1+ \sum_{n=1}^\infty v_{2n} \cos\left(2n \phi \right) + \sum_{n=1}^\infty w^{(h)}_{2n} \sin(2n\phi) \, . 
\end{align}
Several comments are in order. First of all, in the case of an isotropic medium, all Fourier coefficients except for $v_0$ vanish. For an anisotropic medium, both $v_n$ and $w_n^{(h)}$ vanish for odd $n$ due to discrete symmetry $\phi \rightarrow \phi - \pi$ of the distribution. Further, as a consequence of the $\phi \rightarrow -\phi$ antisymmetry of the helicity dependent term, this dependence can only enter via the odd-parity terms $w_n^{(h)}$ in the series. Note that, trivially from Eq.~\eqref{massive_zaligned_final}, one has $w_n^{(h)} \propto h$. Interestingly, to first order in anisotropy, the only harmonics that appear are $v_2$ (elliptic-type deformation) and $w_2^{(h)}$. Hence, in Fig.~\ref{fig:massless_harm}, we show the evolution of these two leading harmonics for general finite $\zeta$ and for massless quarks. First, when comparing the $v_2$ (left) and $w_2^{(+1)}$ (right) harmonics in absolute value, we get a possible explanation for the near symmetry around $\pi/2$ in the plots in Fig.~\ref{azimuthal_plot_phi} for the smallest anisotropies, since this indicates that $\cos(2\phi)$ (an elliptic-type deformation) dominates the $\sin(2\phi)$ term. Looking at the left panel, we see that $v_2$ monotonically increases with the value of the anisotropy parameter, as one should expect. Then, one can also observe that for softer splittings the behaviour with $z$ is non-monotonic and overall quite mild. The evolution of the $w_2$ harmonic (right panel) is non-monotonic with both $\zeta$ and $z$, displaying maxima for intermeadiate values of both quantities. The former is surprising, while the latter is a consequence of the fact that for $z=0,\, 0.5$ there is no helicity dependence.

\begin{figure}
\centering
\begin{subfigure}{0.49\textwidth}
         \centering
         \includegraphics[width=\textwidth]{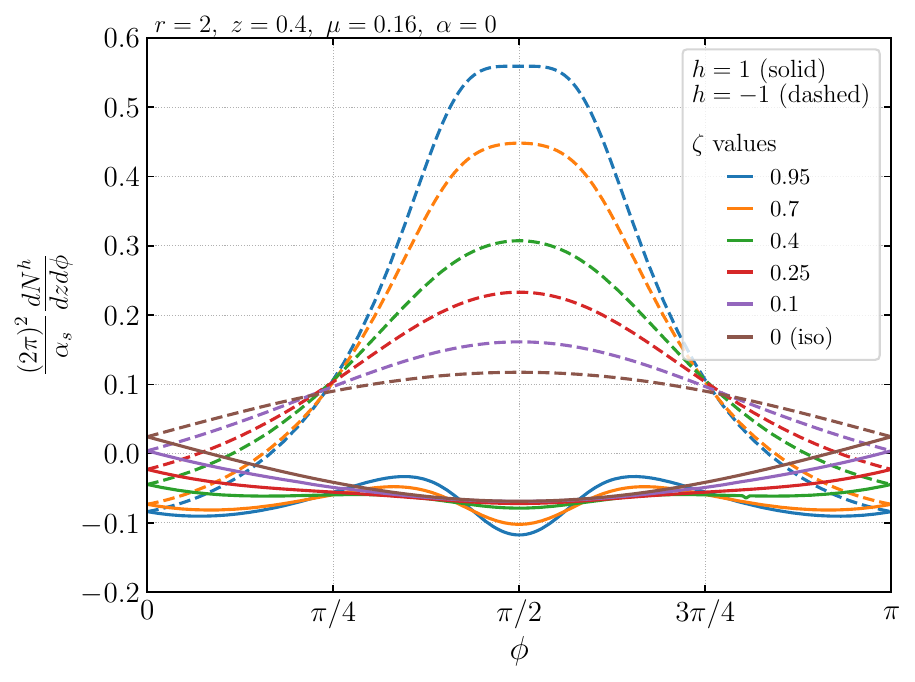}
     \end{subfigure}
     \begin{subfigure}{0.49\textwidth}
         \centering
         \includegraphics[width=\textwidth]{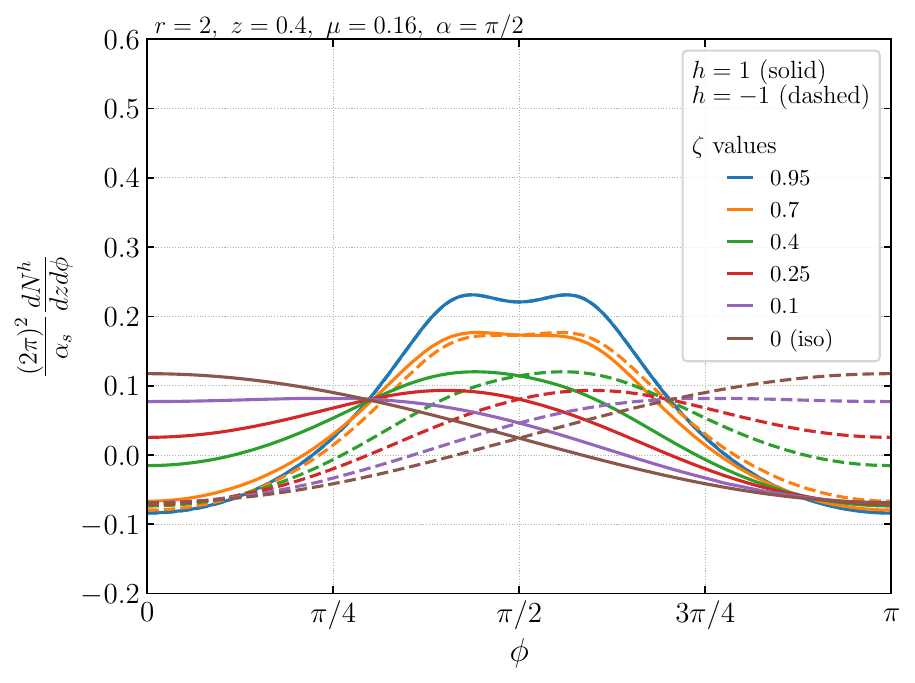}
     \end{subfigure}
\caption{Evolution of the distribution in Eq.~\eqref{massive_tvaligned_final} as a function of $\phi$ for different values of the anisotropy parameter $\zeta$, for $\alpha=0$ (\textbf{left}, spin quantization axis along $x$) and $\alpha = \pi/2$ (\textbf{right}, spin quantization axis along $y$), for fixed $r=2$, $z=0.4$ and $\mu= 0.16$. The vacuum piece is subtracted.}
\label{tvPol_1}      
\end{figure}

\textbf{Transverse polarization for a massive antenna:} In order to tap the interplay between spin and the anisotropy the $q\bar q$ antenna experiences in the transverse plane, it is interesting to consider spin projected along some axis on that plane. One can do so by simply changing the vacuum QCD vertex in the calculation. In this case, the distribution reads
\begin{align}\label{massive_tvaligned_final}
    & \frac{dN^{hh' }}{dzd\phi} = 
    \delta_{hh'}\frac{\alpha_s}{4(2 \pi)^2}\Re  \int_{X_v^+} e^{-i\frac{m^2}{2\tilde{q}_0^+}\Delta t} \frac{\sqrt{c_{1x}}\sqrt{c_{1y}}}{\sqrt{c_{3x}}\sqrt{c_{3y}}}\nn
    & \left(\frac{(c_{1y}c_{2y}c_{3x}-c_{1x}c_{2x}c_{3y})(c_{3y}\cos^2{\phi}-c_{3x}\sin^2{\phi})}{\left(c_{3y}\cos^2{\phi}+c_{3x}\sin^2{\phi}\right)^2} +\left(\frac{m^2}{2\tilde{q}_0^+} + c_4\right)\frac{c_{3x}c_{3y}}{c_{3y}\cos^2{\phi}+c_{3x}\sin^2{\phi}} - \right.\nn
    & \left. -\frac{(1-i)m\sqrt{\pi}}{2\sqrt{\tilde{q}_0^+}}h\sqrt{\frac{c_{3x}c_{3y}}{c_{3y}\cos^2\phi+c_{3x}\sin^2\phi}}\frac{(2ic_{1x}-c_{2x})c_{3y}\cos\phi\sin\alpha - (2ic_{1y}-c_{2y})c_{3x}\sin\phi\cos\alpha}{\left(c_{3y}\cos^2{\phi}+c_{3x}\sin^2{\phi}\right)}\right) \, ,
\end{align}
where $\alpha$ is the angle that the spin quantization axis makes with the $x$-axis, i.e., $\alpha=0,\pi/2$ corresponds to measuring spin along the $x$ and $y$ axis, respectively. Note that we focus only on the equal spins term ($\delta_{hh'}$) in the vertex. It is clear from the previous expression and looking at the $c_{i(x,y)}$ definitions in Eq.~\eqref{eq:ci_definition}, that when there is a very large anisotropy, $\hat q_y \gg \hat q_x$ ($\zeta \rightarrow 1$), the outgoing $q\bar q$ pair is unpolarized along the $y$ direction ($\alpha = \pi/2$). However, when measuring spin along the $x$-axis ($\alpha=0$), the resulting distribution is polarized. This effect is solely driven by the mass dependent term. This observation can be confirmed in Fig.~\ref{tvPol_1}, where we plot Eq.~\eqref{massive_tvaligned_final} for several values of $\zeta$, fixing $z=0.4$ and $\mu = 0.16$, as a function of $\phi$. 
Indeed, for very large anisotropies like $\zeta=0.95$ and $0.7$, one can clearly observe that when $\alpha=\pi/2$ (right panel) the result is nearly independent of $h$ and thus the $q\bar q$ pair is very weakly polarized along the $y$-axis. For $\alpha = 0$ (left panel), the two curves are significantly different around $\phi=\pi/2$, giving rise to a significant polarization along the $x$-axis. 
Even for intermediate values of anisotropy, for instance $\zeta = 0.4$ (the green curve), the $h$ dependence, i.e., degree of polarization, is  larger along $x$ than along the $y$. One should note that in the isotropic case (brown curve in all plots) the $q\bar q$ pair also obtains polarization, but in this case the modulation is trivially given as a function of $\cos(\phi-\alpha)$ (see Eq.~\eqref{massive_tvaligned_final}).

\section{Conclusions}
In this work we have studied $q\bar q$ pair production from a gluon in the presence of an anisotropic, but static and of fixed size, background QCD medium. We introduced anisotropic effects by allowing the jet quenching parameter to differ in two orthogonal directions in the transverse plane with respect to the jet axis, leading to unique angular modulations that couple to the helicity/spin of the final states. We proposed that these modulations can be explored using Fourier azimuthal harmonics inside
jets and through spin polarization measurements in different directions. Since the direction of anisotropy is straightforward to identify experimentally, this offers a new and appealing  window to study directional effects inside jets in heavy ions collisions.

~\\
\textit{\textbf{Acknowledgements:}} We wish to thank A. V. Sadofyev and R. Szafron for helpful discussions. This work is supported by the European Research Council project ERC-2018-ADG-835105 YoctoLHC; by Maria de Maeztu excellence unit grant CEX2023-001318-M and project PID2020-119632GB-I00 funded by MICIU/AEI/10.13039/501100011033; by ERDF/EU; and by FCT, Portugal, under project CERN/FIS-PAR/0032/2021. It has received funding from Xunta de Galicia (CIGUS Network of Research Centres); and from the European Union’s Horizon 2020 research and innovation programme under grant agreement No. 824093. JMS was supported by FCT under contract PRT/BD/152262/2021.

%

\bibliography{refs.bib}

%

\end{document}